# Order recognition by Schubert polynomials generated by optical near-field statistics via nanometre-scale photochromism

Kazuharu Uchiyama[1,*], Sota Nakajima[2], Hirotsugu Suzui[2], Nicolas Chauvet[2], Hayato Saigo[3], Ryoichi Horisaki[2], Kingo Uchida[4], Makoto Naruse[2,*], and Hirokazu Hori[1]

[1] *University of Yamanashi, 4-3-11 Takeda, Kofu, Yamanashi 400-8511, Japan*

[2] *Department of Information Physics and Computing, Graduate School of Information Science and Technology, The University of Tokyo, 7-3-1 Bunkyo-ku, Tokyo 113-8656, Japan*

[3] *Nagahama Institute of Bio-Science and Technology, 1266 Tamura, Nagahama, Shiga 526-0829, Japan*

[4] *Ryukoku University, 1-5 Yokotani, Oe-cho, Seta, Otsu, Shiga 520-2194, Japan*

\* Corresponding author: kuchiyama@yamanashi.ac.jp, makoto_naruse@ipc.i.u-tokyo.ac.jp

**Abstract**

We have previously observed an irregular spatial distribution of photon transmission through a photochromic crystal photoisomerized by a local optical near-field excitation, manifesting complex branching processes via the interplay of deformation of the material and near-field photon transfer therein. Furthermore, by combining such naturally constructed complex photon transmission with a simple photon detection protocol, Schubert polynomials, the foundation of versatile permutation operations in mathematics, have been generated. In this study, we demonstrate an order recognition algorithm inspired by Schubert calculus using optical near-field statistics via nanometre-scale photochromism. More specifically, by utilizing Schubert polynomials generated via optical near-field patterns, we show that the order of slot machines with initially unknown reward probability is successfully recognized. We emphasize that, unlike conventional algorithms in the literature, the proposed principle does not estimate the reward probabilities. Instead, it exploits the inversion relations contained in the Schubert polynomials. To quantitatively evaluate the impact of the Schubert polynomials generated from an optical near-field pattern, order recognition performances are compared with uniformly distributed and spatially strongly skewed probability distributions, where the optical near-field pattern outperforms the others. We found that the number of singularities contained in Schubert polynomials and that of the given problem or considered environment exhibits a clear correspondence, indicating that superior order recognition performances may be attained if the singularity of the given problem is presupposed. This study paves a new way toward nanophotonic intelligent devices and systems by the interplay of complex natural processes and mathematical insights gained by Schubert calculus.

## 1. Introduction

Nanophotonics has been extensively studied not only for energy [1], lighting [2], or sensing applications [3], but also in view of information physics and computing [4–8] to benefit from light-matter interactions in the subwavelength scale. Optical near-field interactions play a crucial role between nanoscale materials [9]. Indeed, nanoscale logic devices have been theoretically and experimentally demonstrated based on optical excitation transfer via optical near-field interactions [10]. Besides, the examination of metamaterials for computing is actively discussed [6]. Furthermore, higher-order intelligent functionalities such as solving satisfiability problems [11] and decision-making problems are investigated [12] using optical near-field interactions.

On the other hand, it is noteworthy that extremely high precision nano-fabrication technologies are indispensable to realize such nanophotonic devices [13]. However, the technical difficulties of ultrafine individual control of the size and the position of nanomaterials, such as quantum dots, are still not easy to resolve to date; this is one of the strong motivations for the development of self-assembly techniques [14, 15]. Furthermore, memory function is another critical factor in view of constructing computing functionality.

Considering these issues, we focus our attention on photochromic materials [16], which exhibit optically induced reversible transformations between transparent or colourless open-ring isomer (1,2-bis(2,4-dimethyl-5-phenyl-3-thienyl)-3,3,4,4,5,5-hexafluoro-1-cyclopentene, **1o**) and opaque or blue-coloured closed-ring isomer (**1c**) (figure 1(a)). Visible light irradiation induces the isomerization from **1c** to **1o** while ultraviolet (UV) does its reverse. Photoisomerization takes the role of information memorization.

Furthermore, it should be emphasized that autonomous pattern formation should be realized in forming nanometre-scale ultrafine structure instead of pixel-wise addressing as in conventional optical memory. To fully benefit from subwavelength-scale photonics, we utilized the interplay involving optical near-fields and locally induced photoisomerization of photochromic materials [17].

In [17], we demonstrated that a local optical near-field excitation on the surface of a photochromic crystal yields local photoisomerization on the scale of tens of nanometres. Initially, the entire material is in a coloured state (**1c**). By an optical near-field excitation with a visible wavelength, fixed at a spatial position, the material is locally isomerized to a transparent **1o** state. The locally induced transparency guides the incoming optical near-field into a neighbouring area in the subwavelength regime. In such a way, the local photoisomerization to the transparent state induces a succeeding chain of local transparency, including bifurcations, leading to the formation of a complex pattern of transmitted photons on the opposite surface of the crystal (figure 1(b)).

Once the nanoscale transparency paths are formed, an incoming near-field excitation is transferred to the opposite side and scattered by the probe tip located at a particular position (figure 1(c)). Indeed, figure 1(d) shows the experimentally observed photon count statistics in the 2 μm × 2 μm regime taken by a scanning near-field optical microscope with a spatial resolution of about 50 nm. We observe non-uniform distributions stemming from the complex structure formed in the photochromic materials [17].

What should be remarked is that, although the photon count pattern looks complicated, the internal photoisomerization is triggered by the local excitation from a single fixed position. Therefore, the photon statistics is not random but contain certain spatial correlations. The origin of the complexity stems from the interplay between local photoisomerization by an optical near-field and its subsequent anisotropic deformation of molecular size, leading to anisotropic mechanical strain [17]. Such a structure is, metaphorically speaking, as if chaos provides complex profiles originating from simple common deterministic nonlinear dynamics. Indeed, by exploiting these characteristics, we have previously successfully demonstrated the generation of Schubert polynomials using the photon count statistics combined with a simple photon detection protocol [18];

we will review the methods in section 2.

However, what was demonstrated by [18] was just the production of Schubert polynomials; its mathematical properties, which are formulated as Schubert calculus, are not fully utilized. Furthermore, quantitative comparisons to other random numbers, such as uniformly distributed pseudorandom numbers, are not examined. In this study, we demonstrate order recognition using Schubert polynomials generated by the experimentally observed optical near-field pattern.

Here, the issue of order recognition concerns the following situation. There are $N$ slot machines whose reward probabilities are given by $P_i$ ($i = 1, \ldots, N$). The task of order recognition means to realize the ranking of all $P_i$ in an ascending or descending order. In this paper, we propose an order recognition strategy based on Schubert calculus. What is remarkable in the proposed strategy is that it never calculates the estimated reward probability $\hat{P}_i$ while estimating the order; instead, the proposed method highlights the relative relation of two choices, which is determined by an inverse relationship in a Schubert polynomial. The details of the principle are introduced in section 3. Furthermore, we quantitatively analyze Schubert polynomials and their order recognition capabilities depending on the property of the source of the randomness. We show that the number of singularities contained in the generated Schubert polynomials is clearly correlated with the order recognition capability.

This paper is organized as follows. In section 2, we review the optical near-field statistics observed by a double-probe optical near-field microscope [17]. Then we review the generations of Schubert polynomials based on near-field observation coupled with a simple photon detection protocol [18]. Section 3 shows the order recognition strategy based on Schubert calculus, including the demonstrations using the experimentally observed optical near-field pattern addressed in section 2. We also quantitatively compare order recognition ability to other random number distributions, such as uniformly distributed random numbers. Section 4 concludes the paper.

## 2. Preparation

### 2.1. Observation of complex optical near-field transmission pattern on photochromic crystal

We first review important previous results, which are the foundation of the present study. In this section, we summarize the complex photon path formation in photochromic materials [17]. As introduced in section 1, the material was a photochromic single crystal of diarylethene with the molecular structures shown in figure 1a [19, 20]. Upon UV irradiation, the molecule is isomerized from the transparent isomer (**1o**) into the blue coloured isomer (**1c**) while maintaining the crystal structure [21, 22]. Upon visible light irradiation, the material was converted back into colourless or transparent isomers. The black and blue curves in figure 1(a) show the absorption spectra of isomers of **1o** and **1c**, respectively. It should be noted that the molecular lengths between two carbons of 4- and 4''- positions of phenyl rings of **1c** and **1o** used in the study are 1.39 nm and 1.41 nm, respectively, and the thickness of the molecule of **1c** and **1o** are 0.39 nm and 0.49 nm, respectively [22], meaning that structural deformations accompany the photoisomerization.

The local optical excitation and its near-field optical transmission characterization were conducted using a double-probe scanning near-field optical microscope (Unisoku, USM-1300S), which has a metallic probe on one side and an apertured optical fibre probe on the other side (figure 1(c)). With the sharpened metallic probe tip located near the surface of the diarylethene crystal with a thickness of 100 μm, local optical excitation was induced by the light at the wavelength of 532 nm. The optical near-field transmission was measured with the optical fibre probe with the spatial resolution of approximately 50 nm, which was connected to a single-photon counting apparatus. The two-dimensional scanning area was 2 μm × 2 μm with a resolution of 256 pixels × 256 pixels.

Figure 1(d) shows the observed near-field photon count distribution obtained after adopting a two-dimensional Gaussian filter with a standard deviation of 6 pixels (approximately 50 nm), corresponding to the resolution limit of the system. We observe complex sub-wavelength structures with a representative scale of 100–200 nm. In particular, the local excitation by the metallic tip did not uniformly propagate through the material. Such a complex pattern generation has been attributed to the balance between the mechanical deformation of the photochromic material and photoisomerization [17, 20]; namely, local photoisomerization induces anisotropic deformation of molecular size, leading to anisotropic mechanical strain [23, 24], which induces subsequent photoisomerization in a non-uniform manner in the surrounding material.

In other words, a chain of spontaneous symmetry breakings was generated in the local photoisomerization involving branching and selection; the near-field-excited diarylethene was locally changed to be transparent. The generated transparent area accompanies an anisotropic mechanical strain in the adjacent area, leading to a suppressed or enhanced photoisomerization to the subsequent photons. That is, an anisotropic strain field from adjacent locally photoisomerized areas provides an anisotropic spatial distribution of the quantum yield of photoisomerization within the crystal [17]. Such chains of photoisomerization and strain-field formation lead to nanometre-scale transparent paths, leading to versatile pattern generation observed in figure 1(d). After transparent path generation, the two-dimensional near-field observation was conducted with the input light by the metallic tip in a single photon level, with photon energies lower than the limit to induce further photoisomerization. Diarylethene, used in this study, does not exhibit thermal isomerization from the closed- to open-ring isomer or vice versa [25]; therefore, the constructed nanometre-scale pattern is preserved until complete resetting via strong UV or visible light irradiation.

### 2.2. Generation of Schubert matrix

The single photon travels through the generated complex transparent paths and is finally transmitted from a particular position on the opposite side. The spatial position of the photon detection differs photon by photon, and figure 1(d) represents its statistics. Uchiyama *et al.* gained new insights in their study in [18] that even though the spatial positions of the photon observations were versatile, it was actually constrained by the transparent paths formed in the diarylethene crystal, which were originally specified by a fixed local excitation from the metallic probe. In order to transform the spatial near-field photon statistics into versatile information in a coherent way, Uchiyama *et al.* incorporated the following simple photon detection protocol to generate Schubert polynomials, which are equivalent to what they call Schubert matrices, as introduced in the following and used in this paper.

#### 2.2.1. Schubert matrix

In a Schubert matrix, there is only a single element of 1 for each and all rows and columns while all the other elements are 0. In other words, a Schubert matrix represents a permutation. A Schubert matrix has a one-to-one correspondence with a Schubert polynomial through divided difference operators [26]. See appendix A.1 for details of the definition of Schubert polynomials.

**2.2.2. Basic notions: diagram, inversion, Young tableau, singularity**

Here we introduce some basic notions to characterize Schubert matrices: *diagram*, *inversion number*, and *singularity*. Here, an example Schubert matrix is a 4 × 4 matrix given by

$$S = \begin{pmatrix} 0 & 1 & 0 & 0 \\ 0 & 0 & 0 & 1 \\ 0 & 0 & 1 & 0 \\ 1 & 0 & 0 & 0 \end{pmatrix}, \quad (1)$$

which is illustrated in figure 2(a) with yellow and green elements denoting 1s and 0s, respectively. Herein, we mark elements that are located on the right and lower sides of the yellow elements, which are marked by the thick black lines in figure 2(b). Therein, the non-marked elements are called the *diagram* of the Schubert matrix, which in this case is given by

$$\begin{pmatrix} * & 0 & 0 & 0 \\ * & 0 & * & 0 \\ * & 0 & 0 & 0 \\ 0 & 0 & 0 & 0 \end{pmatrix}. \quad (2)$$

The number of ∗s in the diagram is called the *inversion number*, which is 4 in the case of the present example. An array of ∗s located at the upper left corner of the matrix is called Young tableau [27]. Further, in the diagram, let connect ∗s into blocks if they are connected by any of their four neighbours: up, down, left, or right. We then count the number of blocks of connected ∗s, which is two in the case of equation (2). Here, the number of *singularities* is defined by the number of such blocks without counting the Young tableau. In the case of equation (2), there are two blocks of ∗, but one of them is the Young tableau. Therefore, the singularity is 1.

**2.2.3. How to generate a Schubert matrix**

In [18], the following protocol was demonstrated to generate a Schubert matrix $S$. Here, we introduce the procedure along with the case of generating a Schubert matrix with size 4 × 4. As shown in section 2.1, a single photon is observed at a spatial position with the corresponding probability distribution (figure 3(a)). The details of the conversion from the experimentally observed photon statistics to selection probability are described in the *Methods* section of [18].

(1) The probability distribution specified by the near-field observation is rescaled or coarse-grained into a $N \times N$ grid. The probability at each point of the grid is denoted by $P_{i,j}$ ($i = 1,..., N, j = 1,..., N$) hereafter, where $N$ is a natural number. In the example, $N$ is equal to 4 (figure 3(a)). In the physical system shown in figure 1(c), when sending a single photon, we observe it at one of the grid points. In this study, by using uniformly distributed pseudorandom numbers generated in a computer and pondered by the measured probabilities, we emulate the output photon observations.

(2) We count the number of photons detected at each grid element. Let the photon count at the grid point $(i, j)$ be denoted by $C_{i,j}$ ($i = 1,..., N, j = 1,..., N$). When the photon count at a certain position $C_{m,n}$ becomes larger than a threshold value denoted by $P_T$, we determine that the ($m$, $n$)-element of the Schubert matrix is given by 1; that is, $S_{m,n} = 1$. In the case of the example in figure 3(b), $S_{3,3} = 1$.

(3) Accordingly, the photon detection probability distribution $P_{i,j}$ is updated for the rest of the loop. First, photon detection at the $m$-th row and the $n$-th column is configured to be zero, which are denoted by X marks in figure 3(c). That is, $P_{3,j} = 0$ ($j = 1,..., 4$) and $P_{i,3} = 0$ ($i = 1,..., 4$). This reconfiguration is based on the definition of the Schubert matrix. Second, the probability distribution $P_{i,j}$ is renormalized so that $\sum_{i,j} P_{i,j} = 1$ holds.

(4) Similarly to Step (2) above, when $C_{m',n'}$ becomes larger than $P_T$, we determine the ($m'$, $n'$)-element of the Schubert matrix to be 1; that is, $S_{m',n'} = 1$. In the case of the example in figure 3(d), $S_{2,4} = 1$.

(5) We repeat such Steps of (2) and (3) two more times (figures 3(e), 3(f), 3(g), 3(h)); $S_{1,2} = 1$ is determined as shown in figure 3(f), followed by $S_{4,1} = 1$ in figure 3(h).

One minor remark here is that the final determination of $S_{1,2} = 1$ is immediately given after the previous step because only a single grid point has a non-zero photon observation probability. The resultant Schubert polynomial is $x_1^2 x_2 x_3 + x_1 x_2^2 x_3$; see appendix A.1 for details about Schubert polynomials [26].

The total number of different Schubert matrices of size $N \times N$ is given by $N!$, meaning a factorial order increase with the size of the observation elements. With $N = 32$, the total number is more than $10^{35}$. Such extremely diverse pattern generation abilities originating from the same substrate via optical near-field processes were the central topics studied in [18].

The photon number threshold $P_T$ plays a central role here. A smaller $P_T$ value leads to diverse choices of 1s in the generated Schubert matrix. Conversely, a larger $P_T$ value indicates that the element having the maximum photon detection probability most likely yields the value of 1 in the resulting Schubert matrix. In other words, the resultant Schubert matrices are more weakly and more strongly correlated with the source optical near-field pattern with smaller and larger $P_T$ values, respectively [18]. These are called *soft* and *hard* correspondence between near-field patterns and Schubert matrices in [18].

## 3. Order recognition
### 3.1. Principle
Section 2 reviews important backgrounds and fundamentals needed for the present study, concluded with the generation of Schubert matrices based on near-field photon statistics from a photochromic material. The primary focus of this paper is to utilize these Schubert matrices for information processing functions; more precisely, our interest is to recognize the order or ranking of slot machines with initially unknown reward probability as introduced in section 1.

### 3.2. Order recognition as the revision of *name* and *value*
The key is to exploit the inversion relation presented by a Schubert matrix; here, we introduce it with a 4 × 4 matrix, which is schematically illustrated in figure 4(a). If the column number of the element with the value of 1 in the *i*-th row is smaller than that in the (*i* + 1)-th row, we define that these two elements are in a relation of *inversion*. Furthermore, this inversion is related to the concept of *singularity* introduced in section 2.2.2. The diagram of the Schubert matrix (figure 4(a)) is shown in figure 4(b).

Conversely, when the Schubert matrix is given by an anti-diagonal matrix (figure 4(c)), there are no inverse relations therein. Also, the diagram of an anti-diagonal matrix is shown in figure 4(d) where the upper triangular elements, the Young tableau, only contribute to forming the inversion number. The inversion number is maximized. There is no singularity. The Schubert polynomial is represented by a single term $x_1^3 x_2^2 x_3$.

As we observe above, inversion relations and singularity indicate characteristic properties. The idea of the proposed ordering strategy based on Schubert calculus begins with associating information to the vertical (row) and the horizontal (column) axis of Schubert matrices.

Assume that there are four slot machines $M_1$, $M_2$, $M_3$, $M_4$, and their reward probabilities are given by $(P_1, P_2, P_3, P_4) = (0.4, 0.2, 0.8, 0.6)$, which are unknown to us. In this case, the ground truth order of the slot machines' reward probability is $M_3$, $M_4$, $M_1$, $M_2$ in descending order. We want to recognize this ground truth order by playing the slot machines.

The first critical point of the proposed order recognition strategy is to associate the vertical (row) and the horizontal (column) axes of a matrix with *name* and *value*, respectively. More specifically, the name of slot machines is bound with the rows while the upper rows indicate a higher-ranked machine. The lower rows mean a lower-ranked machine. Meanwhile, the columns correspond to the reward probability, with the right column being high reward probability. These associations are illustrated in figure 4(e). In the pseudo-code shown below, these associations are represented by line 5.

Suppose that the elements along the name axis are perfectly arranged in the order of the ground truth. Also, the ground truth reward probabilities are perfectly arranged in ascending order. Let the element specified by the name and its value be 1. Then, the matrix results in an anti-diagonal matrix, as shown in figure 4(c). Here, there are no inverse relations and no singularities. The inversion number is its maximum. That is, perfect order recognition is represented by the anti-diagonal matrix. Therefore, transforming the initially unsure order regarding both name and value axes to an anti-diagonal matrix corresponds to accomplishing the correct order recognition. The question now is how to realize such transformations.

### 3.3. Order recognition as the revision of *name* and *value*

We consider that the Schubert matrices *propose* which machines should be checked to determine the correct order. In this study, a pair of machines, namely, two machines, are selected in a single slot machine play. This is similar to the multi-armed bandit (MAB) problem with multiple plays [28], although the aim of solving MAB problem and the present study are fundamentally different. The purpose of solving MAB problems is usually minimizing regret, whereas the aim of the study here is to recognize the order.

What is important is that the machine selection is determined by the inverse relation in the Schubert matrix. In the case of the example shown in figure 4(a), there is one inverse relation indicated by a solid blue arrow.

We pay our attention to the inverse relation; in the case of figure 4(a); the inversion regards the first and the second rows. We assume that the name of the first and the second row is machine A and machine B, respectively. Based on the definition of the vertical and horizontal axes of the matrix shown in figure 4(e), the relation between machine A and machine B is *contradictory* (figure 4(f)). According to the name axis (vertical), machine A is considered to be a higher reward machine. However, the value axis (horizontal) indicates that the value of machine B is greater than machine A. That is, the relation is contradictory.

To resolve such a contradiction, we do select these two machines and check the betting results.

[CASE 1] When the reward dispensed by machine A is greater than that of machine B:
> The order of the name axis is correct; machine A is above machine B. On the other hand, the value axis is *wrong*; the value of machine A should be on the right-hand side of machine B. Therefore, we flip the association of the value axis, leading to a situation depicted in figure 4(g).

[CASE 2] When the reward dispensed by machine B is greater than that of machine A:
> The order of the value axis is correct; the value of machine B is on the right-hand side of machine A. On the other hand, the name axis is *wrong*; machine B should be above machine A. Therefore, we flip the association of the name axis, leading to a situation depicted in figure 4(h).

It should be noted that the resulting arrangement is actually the same as shown in figures 4(g) and 4(h) for both cases; however, the reconfiguration of the axes is totally contrasting.

After finishing the reconfiguration by the currently used Schubert matrix, we continue the process by using the next Schubert matrix, which is shown by line 4 in the pseudo-code below (`S ← Schubert matrix`). Furthermore, the row and column of the new Schubert matrix are transformed by the name and value axes to date, which is represented by line 5 in the pseudo-code (`S ← S[name,value]`).

The above-described principle is summarized as a pseudo-code by the following:

**Algorithm:** Order recognition

1    *name* ← 1,…, *N*

```
2   value ← 1,…, N
3   loop
4       S ← Schubert matrix
5       S ← S[name,value]
6       for i in S where the rows i and i + 1 are in an inverse relation, do
7           A = pull(name[i])        % A is the reward from the machine name[i]
8           B = pull(name[i + 1])    % B is the reward from the machine name[i + 1]
9           if A > B then
10              flip(name[i],name[i + 1])
11          else
12              flip(value[name[i]],value[name[i + 1]])
13          end if
14      end for
15  end loop
```

### 3.4. Demonstrations

We demonstrate the order recognition with 16 slot machines ($M_1,..., M_{16}$) with the ground-truth reward probability setting of ($P_1,..., P_{16}$) as shown by the bar graph in figure 5(a,i). The true order of the reward probability in descending order is given by $M_9$, $M_{14}$, $M_8$, $M_{11}$, $M_2$, $M_6$, $M_3$, $M_{13}$, $M_1$, $M_4$, $M_{12}$, $M_7$, $M_5$, $M_{10}$, $M_{16}$, $M_{15}$ as shown in figure 5(a,ii).

Initially, the association of the name axis starts with the linear order corresponding to the index of the machines; that is, $M_1, M_2,..., M_{16}$. We use a total of 10000 Schubert matrices in a consecutive manner. This whole procedure was repeated $R = 100$ trials. Since the reward dispensed by each of the machines is probabilistic, the win-lose event occurs differently among $R$-times trials.

The images in figure 5(b) demonstrate how the order recognition evolves as a function of the number of Schubert matrices used for the ordering, denoted by $t$, implying the time, in the horizontal axis. Here, the order of machine $M_i$ at time $t$ is evaluated by calculating how often the machine $M_i$ is associated with the row number of the Schubert matrices. Meanwhile, we can compute how the value of $M_i$ is evaluated at time $t$ by analyzing how its value is ranked in the horizontal direction.

The two-dimensional images in figure 5(b) show how the order recognition evolves by counting the accumulated incidences of the matrix of (ranking, value), referred to as *ranking-and-value matrix* hereafter, at $t = 100, 200, 500, 1000$. If the order recognition is perfectly completed, the top-ranked machine owns the highest value while the worst-ranked machine has the minimum value. Therefore, the ranking-and-value matrix exhibits higher incidences in the anti-diagonal axis. Indeed, in the initial phase at $t = 100$, the matrix shows a diverse incidence pattern. As time elapses, the high incidence elements are gradually localized along the anti-diagonal axis, indicating successful order recognition.

Figure 5(c) visualizes the results in another way where the vertical axis represents the estimated order while the horizontal axis shows the number of Schubert matrices used or the time dimension. Precisely, the estimated order of machine $M_i$ at time $t$ is calculated by the average of the estimated ranking of $M_i$ by time $t$. Furthermore, such estimated order is averaged over 100 different trials. The 16 curves therein, shown with different colours, represent the time evolution of how machine $M_i$ is ranked. We observe that the estimated order is drastically reconfigured during the first about 500 steps. When the time is approximately 2000, the order recognition is accomplished.

Furthermore, figure 5(d) summarizes the number of correctly order recognized machines as a function of time, which equals 16 when the perfect order recognition is achieved. Indeed, after $t = 2298$, the estimated order agrees with the ground truth ranking. We can also observe that in the very stage at $t = 626$ the order of most of the machines, 14 out of 16, is already recognized.

In the meantime, the error bars of figure 5(c) indicate the standard deviation of the estimated order; when the ground-truth reward probability difference between slot machines is minute, the error bars of associated machines overlap with each other. We observe such a situation regarding $M_{15}$ and $M_{16}$ where the reward probability difference is only approximately 0.02.

However, it should be emphasized that we can automatically grasp such uncertainty by the ranking-and-value matrix. Indeed, we can observe a high incidence square area, or a block matrix, in the lower-left corner of the matrix shown in figure 5(b,iv). This means that the ranking of the 15$^{th}$ and 16$^{th}$ machines are comparable. Indeed, this is a remarkable attribute of the proposed strategy based on Schubert polynomials in the sense that ordering is accomplished while maintaining strong correlations inherent in the problem under study.

Furthermore, as mentioned in section 1, the proposed order recognition strategy relies only on the two choices suggested by Schubert matrices and their resulting relationships. The notions of reward probability and confidence intervals are not involved at all in the present study; this is clearly contrasting to the former order recognition principles [29].

### 3.5. Comparison

Section 3.4 shows the successful demonstration of order recognition in a 16-machine probabilistic reward environment based on experimentally observed near-field photon statistics. Here we discuss the differences of Schubert matrices and their order recognition performances among different random number generator pattern statistics. More specifically, the other source patterns are spatially uniform and spatially strongly skewed probabilities, as shown in figure 6(a) denoted by "uniform" and figure 6(b) denoted by "centre", respectively. The latter one has a larger probability in the square area located at the centre. The detailed setting of the "centre" is described in appendix A.2. The near-field photon observation pattern (figure 1(d)) is simply called "near-field" hereafter.

### 3.5.1. Schubert matrix statistics

We generate Schubert matrices based on near-field, uniform, and centre patterns, respectively, with $P_T$ value of 1. The square, x, and circular marks in figure 6(c,i) evaluate the average value of the inversion number of the generated Schubert matrices from near-field, uniform, and centre patterns, respectively, as a function of the spatial resolution, specified by $N$ introduced in section 2.2.3, ranging from 8 to 24. The detailed setting of Schubert matrix generation is described in appendix A.2. Figures 6(c,ii) and 6(c,iii) summarize the median and standard deviation of the inversion number. From figure 6(c), we do not observe significant source-signal dependencies in these basic statistics about the inversion number.

Similarly, we examine the average (figure 6(d,i)), median (figure 6(d,ii)), and standard deviation (figure 6(d,iii)) of the singularity of the generated Schubert matrices from the near-field, uniform, and centre patterns, respectively. There are also no clear source-signal dependencies regarding singularity.

By contrast, evident source-signal dependencies are observed by taking into account the uniqueness of Schubert matrices. The blue, green, and red curves in figure 6(e,i) examine the frequency of Schubert matrices divided by the unique number of Schubert matrices as a function of their inversion number regarding near-field, uniform, and centre patterns, respectively, when $N$ is given by 8. Whereas the uniform pattern yields an almost identical frequency regardless of the inversion number, the near-field pattern shows a peak frequency around the inversion number of 10. Similar trends are observed when $N$ is given by 9 and 10, as demonstrated in figures 6(e,ii) and 6(e,iii), respectively.

Likewise, the frequency of Schubert matrices divided by the unique number of Schubert matrices exhibits a similar tendency with respect to the number of singularities, as shown in figures 6(f,i) ($N$ = 8), 6(f,ii) ($N$ = 9), and 6(f,iii) ($N$ = 10). While the uniform pattern exhibits similar values for all singularities, the near-field provides a larger frequency as the singularity increases. Remember that the statistics are very similar among these three source patterns. Therefore, the observation in figure 6(f) means that versatile Schubert matrices are generated regarding small singularities via the near-field pattern. In the meantime, by the centre pattern, the generated Schubert matrices are mostly similar to each other when their number of singularities is small.

### 3.5.2. Order recognition performance comparisons

We examine the order recognition performances depending on the source patterns regarding eight slot machine ($M_1$,..., $M_8$) environments. While the demonstrations shown in section 3.3 suppose a probabilistic reward environment, here we assume deterministic, different-valued rewards among (0.2, 0.3, 0.4, 0.5, 0.6, 0.7, 0.8, 0.9). The correspondence between $M_1$,..., $M_8$ and the rewards is randomly configured; we examine $K$ = 8! = 40320 kinds of unique correspondences between the machines and the rewards.

For a given reward environment, consecutive 250-times slot machine plays are conducted. Such plays are repeated for $K$-kinds of different reward environments. The correct order recognition rate (COR) at time step $t$ is defined by the number of times when the order of the reward is correctly recognized at time step $t$ among $K$-kinds of different reward environments divided by $K$. The blue, green, and red curves in figure 7(a) represent the time evolution of COR when the near-field, uniform, and centre patterns are utilized respectively in the source Schubert matrix generation when the photon threshold $P_T$ was given by 1. We observe that the near-field most promptly achieved a higher COR than others. On the contrary, the centre pattern cannot conduct correct order recognition for certain reward environments because COR does not converge to unity.

The order recognition ability also depends on the photon threshold $P_T$. Figure 7(b) characterizes the number of steps when COR exceeds the value of 0.9 as a function of photon threshold $P_T$. The near-field pattern, denoted by the blue curve, achieves smaller steps than others; in particular, the minimum steps was accomplished when $P_T = 8$. In the meantime, the uniform and centre patterns, marked by the green and red curves, respectively, do not show clear $P_T$ dependencies, which is naturally understood from the fact that the source pattern has no spatial dependencies in the uniform pattern and the centre pattern does not provide different spatial inhomogeneity by changing $P_T$.

Finally, we examine order recognition from the viewpoint of singularities. As mentioned at the beginning of this section, there are in total $K = 40320$ assignments of the rewards to the eight slot machines. Each assignment can be represented by a Schubert matrix. Therefore, we can derive the singularities of the given assignment. Figure 7(c) shows the number of assignments, with the number being annotated with the bar graph, as a function of the number of singularities ranging from 0 to 6.

We examine the COR depending on the singularity number of the assignments, ranging from 0 to 5, as shown in figure 7(d). When the singularity number is smaller than or equal to two, the near-field pattern accomplishes the fastest adaptation to COR = 1, or the perfect order recognition, whereas the COR by the centre pattern is limited below unity. Conversely, beyond the singularity number of 3, the COR exhibits comparable performance irrelevant to the source signal patterns. But it is notable that with the singularity number of 5, the centre pattern outperforms the others.

Such trend is contravariant to the trend of the frequency of the unique Schubert matrix analyzed in figure 6(f). That is, the near-field pattern yields versatile Schubert matrices when there are few singularities, and it provides superior ordering performances for slot machine assignments with small singularities. With the centre pattern, on the other hand, the versatility of Schubert matrices is low when there are few singularities, and the order recognition ability is inferior when the singularity number of the problem is small. Overall, as shown in figure 7(c), the main part of all arrangements consists of fewer singularities configurations; therefore, on average, the near-field provides superior COR, as summarized in figure 7(a).

A clear understanding of the underlying mechanism behind the relationship between the singularities of the source Schubert matrices and the order recognition is a very interesting future topic. At this time, we have the following speculations. The abovementioned results indicate that the matching of the versatility of Schubert matrices with the number of singularities of the given problem plays a key role. Presumably, by subjecting a variety of different Schubert matrices that is matched with the singularity number of the given problem, the revision of the ordering proceeds efficiently. If a limited kind of Schubert matrices is being used, the order cannot be correctly recognized due to insufficient comparisons. In connection with such results, Okada *et al*. demonstrated that negative autocorrelation contained in a chaotic time series accelerates solving the two-armed bandit problems in [30], which also manifests benefits from structured randomness for functionality.

In the meantime, if one can know the singularity of the given problem as prior knowledge, the acceleration of the order recognition will be possible by engineering the source pattern that contains appropriate singularities. In the case of the discussion above, for example, if the singularity number is of the given problem is 5 or 6 only, the usage of the centre pattern is beneficial rather than the near-field or the uniform patterns. A deeper understanding of related issues may be another interesting future study.

## 4. Conclusion

We demonstrate an order recognition algorithm based on Schubert polynomials generated via irregular spatial distribution of photon transmission through a photochromic crystal photoisomerized by a local optical near-field excitation. The proposed strategy exploits the inversion relations and the singularities inherent in Schubert polynomials. A successful order recognition is demonstrated regarding 16 different reward probability slot machines. Furthermore, the performance of order recognition is quantitatively compared among the source Schubert matrices generated via near-field, spatially uniform and strongly skewed patterns. We found that the singularity number contained in Schubert matrices and that of the given problems exhibit a clear correlation, suggesting that the order recognition will accelerate if the singularity number of the considered environment is presupposed. This study paves the way toward nanophotonic intelligent devices and systems by the interplay of complex natural processes and mathematical insights gained by Schubert calculus.

**Acknowledgments**

This work was supported in part by the CREST project (JPMJCR17N2) funded by the Japan Science and Technology Agency and Grants-in-Aid for Scientific Research (A) (JP20H00233) and Grant-in-Aid for Challenging Research (Exploratory) (JP21K18710) funded by the Japan Society for the Promotion of Science.

**Appendix A. Schubert polynomials**

**Appendix A.1. Definition of Schubert polynomials**

The Schubert polynomial $\{\mathfrak{S}_\omega\} = \{\mathfrak{S}_\omega(x)\} = \{\mathfrak{S}_\omega(x_1,\cdots,x_{n-1})\}$ [26, 27, 31, 32] about symmetric group $S_n$ is defined through the divided difference operator $\partial_{xy}$ given by

$$\partial_{xy} f = \frac{f(x,y) - f(y,x)}{x - y}. \tag{A.1}$$

Let the polynomial of the permutation with the maximum inversion number be given by

$$\mathfrak{S}_{\omega_0} = x_1^{n-1} x_2^{n-2} \cdots x_{n-1}. \tag{A.2}$$

For a general permutation $\omega$, the polynomial is defined by

$$\mathfrak{S}_\omega = \partial_{\omega^{-1}\omega_0} \mathfrak{S}_{\omega_0}. \tag{A.2}$$

In the case of the four-dimensional symmetric group $S_4$, the permutation with the largest inversion number is $\omega_0 = [4321]$, which is represented by the anti-diagonal matrix shown as in figure 4(c). Its inversion number is six, as shown in figure 4(d). The corresponding Schubert polynomial is $\mathfrak{S}_{\omega_0} = \mathfrak{S}_{4321} = x_1^3 x_2^2 x_3^1$.

The Schubert polynomial of the permutation [4231] is given by the divided difference operation on $\mathfrak{S}_{\omega_0}$ regarding $x_2$ and $x_3$. Hence, $\mathfrak{S}_{4231} = (x_1^3 x_2^2 x_3^1 - x_1^3 x_3^2 x_2^1)/(x_2 - x_3) = x_1^3 x_2 x_3$ is obtained. Similarly, the Schubert polynomial of [2431] is given by the divided difference of $\mathfrak{S}_{4231}$ with respect to $x_1$ and $x_3$. Therefore, $\mathfrak{S}_{2431} = (x_1^3 x_2 x_3 - x_3^3 x_1 x_2)/(x_1 - x_2) = x_1^2 x_2 x_3 + x_1 x_2^2 x_3$ is obtained, which is the Schubert polynomial corresponding to the Schubert matrix shown in figure 4(a). All of the Schubert polynomials, inversion numbers, and singularity of $S_4$ are summarized in table A.1.

**Appendix A.2. Simulation details**

In section 3.5, we generated Schubert matrices from a spatially strongly skewed probability distribution called "centre". Here we describe the details of this pattern. When the size of the matrix under study is $N \times N$, the area of the centre is the set of elements $(i, j)$ taking the integer numbers of $\lceil 7N/16 \rceil \le i \le \lceil 9N/16 \rceil$ and $\lceil 7N/16 \rceil \le j \le \lceil 9N/16 \rceil$. The detection probability of a pixel belonging to the centre is 100 times larger than that of a pixel outside the centre area.

In section 3.5.1, we examined statistics of the Schubert matrices regarding $N = 8$, 9, 10, 11, 12, 16, 20, and 24. For each of these cases, the number of generated Schubert matrices is summarized in table A.2. Since the total number of different Schubert matrices is $N!$, the number of generated matrices should be sufficiently high to investigate the statistical characteristics. However, the computational demands are too high, especially when $N$ is larger than 11, in our computing environment. For these reasons, the number of the generated matrices is limited to $2 \times 10^9$ when $N$ is larger than or equal to 11.

The computing environment is a workstation of HPC Systems (CPU: Xeon Gold 6249R (20 cores, 3.1 GHz), RAM: 384 GB, OS: Ubuntu), and the software platform is Matlab.

**ORCID IDs**

K Uchiyama https://orcid.org/0000-0002-6919-8903
N Chauvet https://orcid.org/0000-0002-6504-1730
H Saigo https://orcid.org/0000-0002-4209-352X
R Horisaki https://orcid.org/0000-0002-2280-5921
K Uchida https://orcid.org/0000-0001-5937-0397
M Naruse https://orcid.org/0000-0001-8982-9824

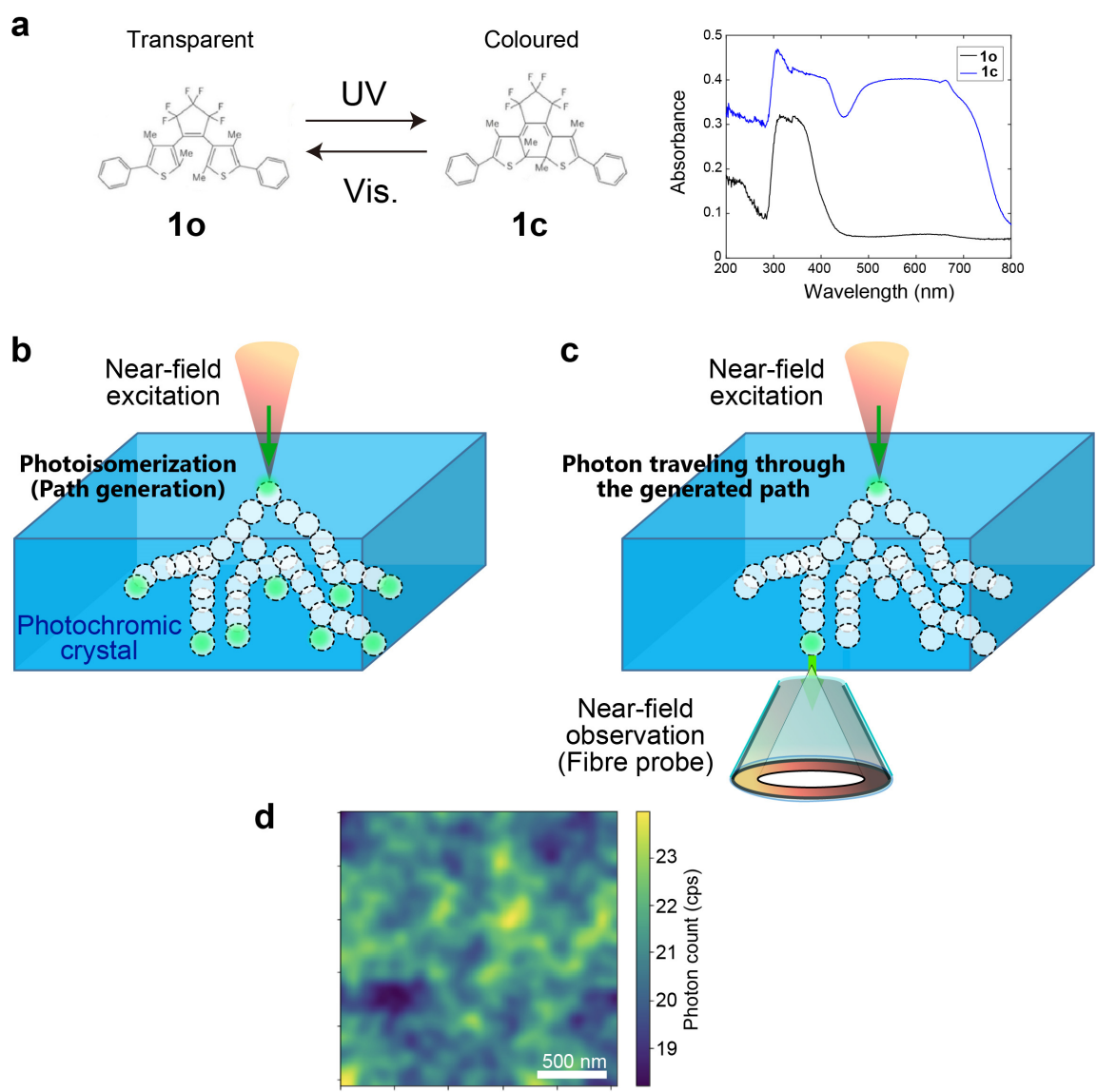

**Figure 1.** Nanoscale photochromism by near-field optics. (a) Transparent and coloured states of diarylethene and their absorption spectra. (b) Local near-field light excitation at the surface of the photochromic material to generate photoisomerization at the nanometre scale. Complex pathways are generated. (c) Once the pattern is formed, the local optical excitation travels through the generated path. (d) The output position of the photons was observed by scanning near-field optical microscopy. Adapted from Uchiyama *et al.*, Sci. Rep. 10, 2710 (2020). Copyright 2020 Author(s), licensed under a Creative Commons Attribution 4.0 License.

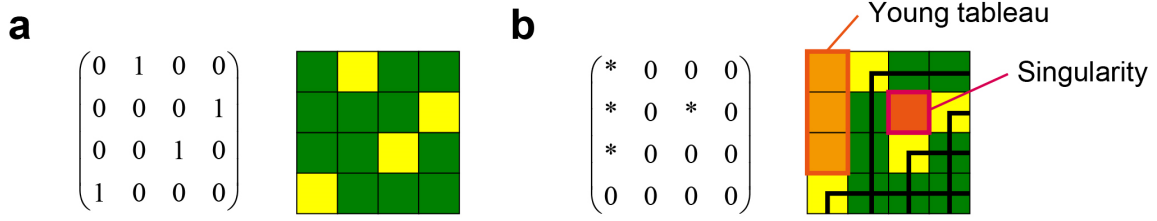

**Figure 2.** Schematic illustration of an example of Schubert matrix. (a) The permutation of [2431] is represented by a matrix *S* in equation (2). (b) The *diagram* of the matrix *S* is illustrated. The upper-left block consisting of ∗s (indicated by orange squares) is called *Young tableau*, whereas isolated ∗ blocks are called *singularities*.

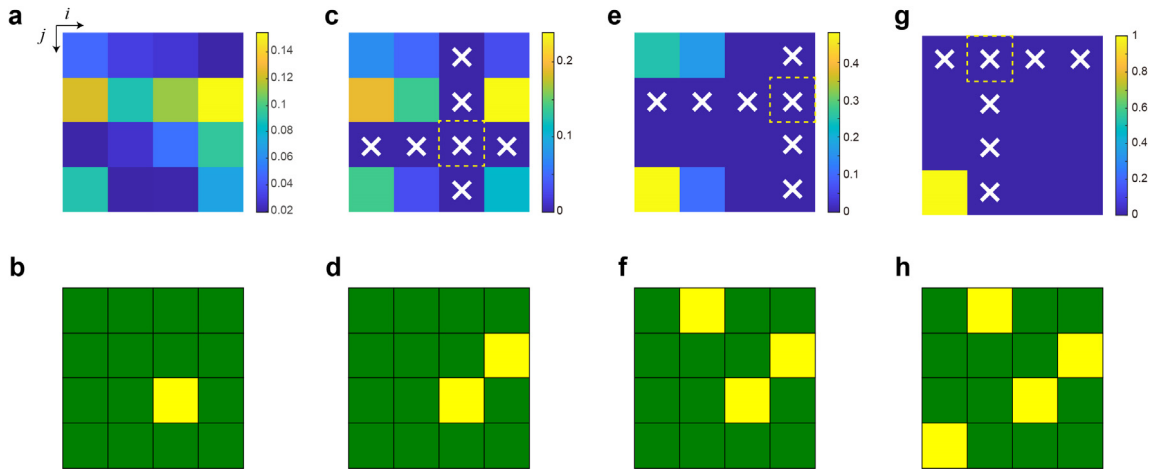

**Figure 3.** How to generate Schubert polynomials based on optical near-field observation. (a) The detection area is divided into $N \times N$ elements. Here $N = 4$. The photon detection probability is spatially irregular. (b) When a photon is detected at an element $(i, j)$, the $(i, j)$-element of Schubert matrix is given by 1. Here, $(i, j) = (3,3)$. (c) The sensitivities of all the elements of the 3$^{rd}$ row and column are turned to be zero. The probabilities are renormalized. (d) Next element of Schubert matrix is determined. (e,f,g,h) Repeat the process of (c) and (d). (h) is the finally obtained Schubert matrix in this example. Adapted from Uchiyama *et al.*, Sci. Rep. 10, 2710 (2020). Copyright 2020 Author(s), licensed under a Creative Commons Attribution 4.0 License.
23

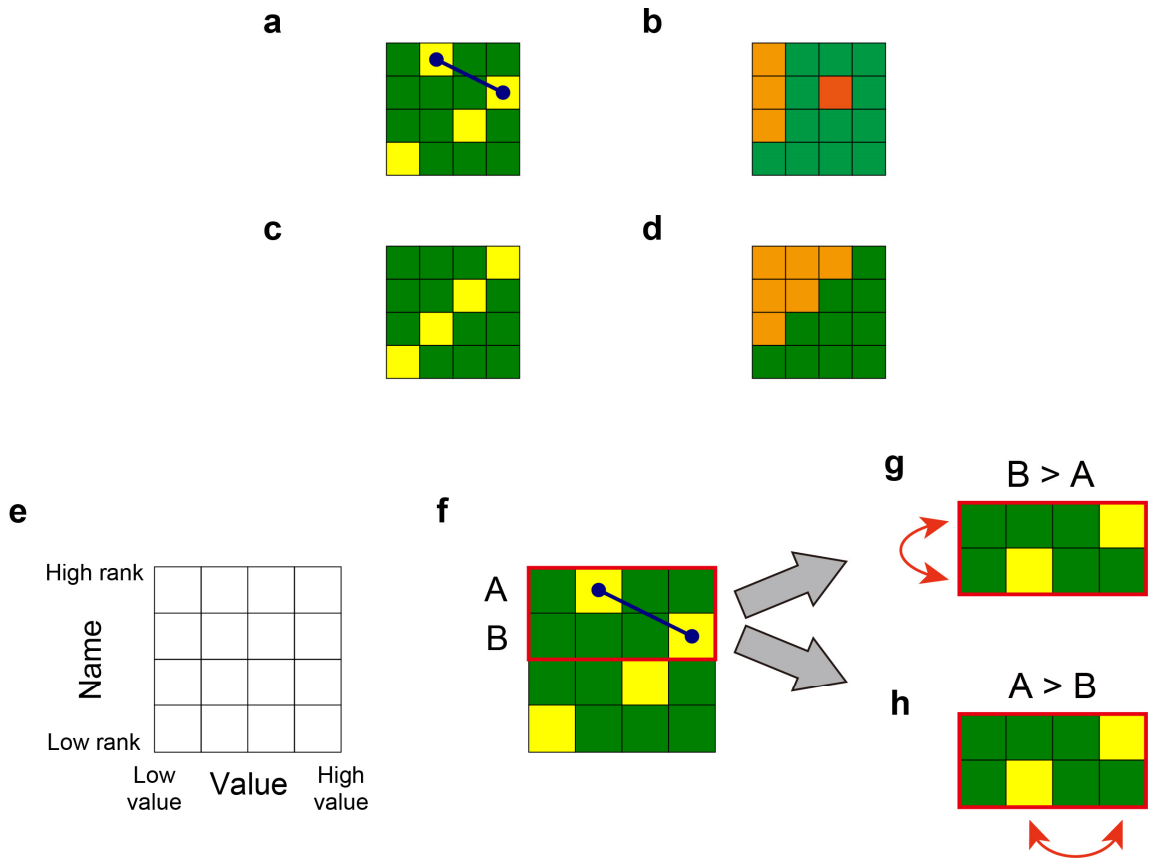

**Figure 4.** Order recognition principle based on Schubert polynomials. (a) Inversion relation (marked by a blue arrow) inherent in the Schubert matrix. (b) Young tableau (orange colour) and a singularity (red colour). (c) Illustration of an anti-diagonal matrix and (d) its diagram contains a Young tableau with the maximum inversion number. (e) In the proposed ordering strategy, a matrix is characterized by the *name* and *value* axes, which respectively correspond to the vertical and horizontal direction. In the name axis, the upper row indicates a high rank. In the value axis, the right column means a high value. (f) Our decision of machine selection is suggested by the Schubert matrix. We pay attention to the inversion relation of the given matrix, which is the 1$^{st}$ and the 2$^{nd}$ row in (f). Let the names of the 1$^{st}$ and 2$^{nd}$ rows be machine A and B, respectively. The inversion relation means a contrary situation among the name and value axes. (g) If the reward dispensed by machine B is greater than A, the association of the name axis was wrong; hence the allocation of the names are swapped. (h) If the reward by machine A is larger than B, the association of the value axis was wrong; hence the allocation of the values is flipped.

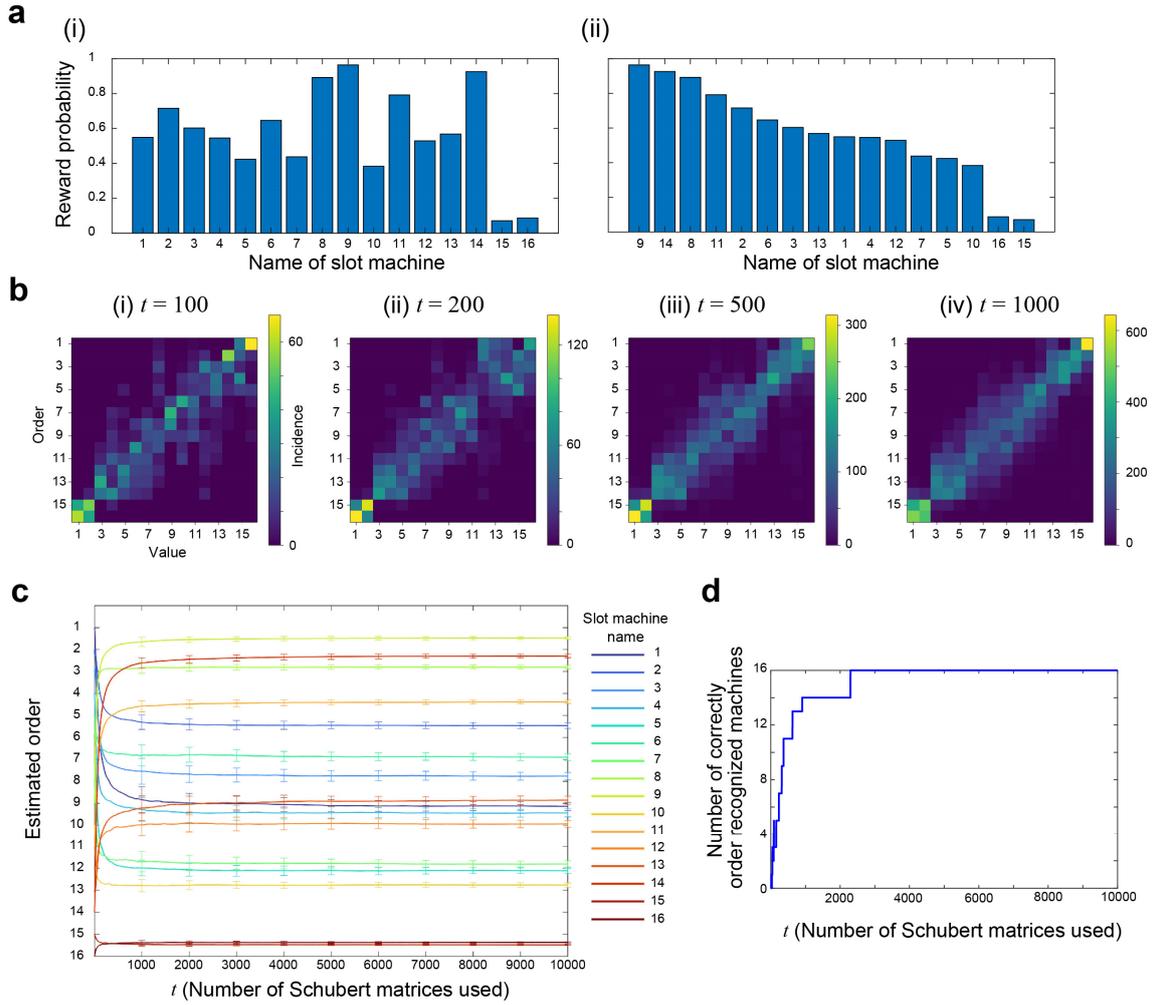

**Figure 5.** Demonstration of order recognition in 16 slot machine environments. (a) (i) Reward probability arrangement, and (ii) its sorted representation. (b) The accumulated incidence pattern of the ranking-and-value matrix until the time step of (i) 100, (ii) 200, (iii) 500, and (iv) 1000. (c) The time evolution of the estimated ranking of each slot machine. (d) The number of correctly ordered machines over time.

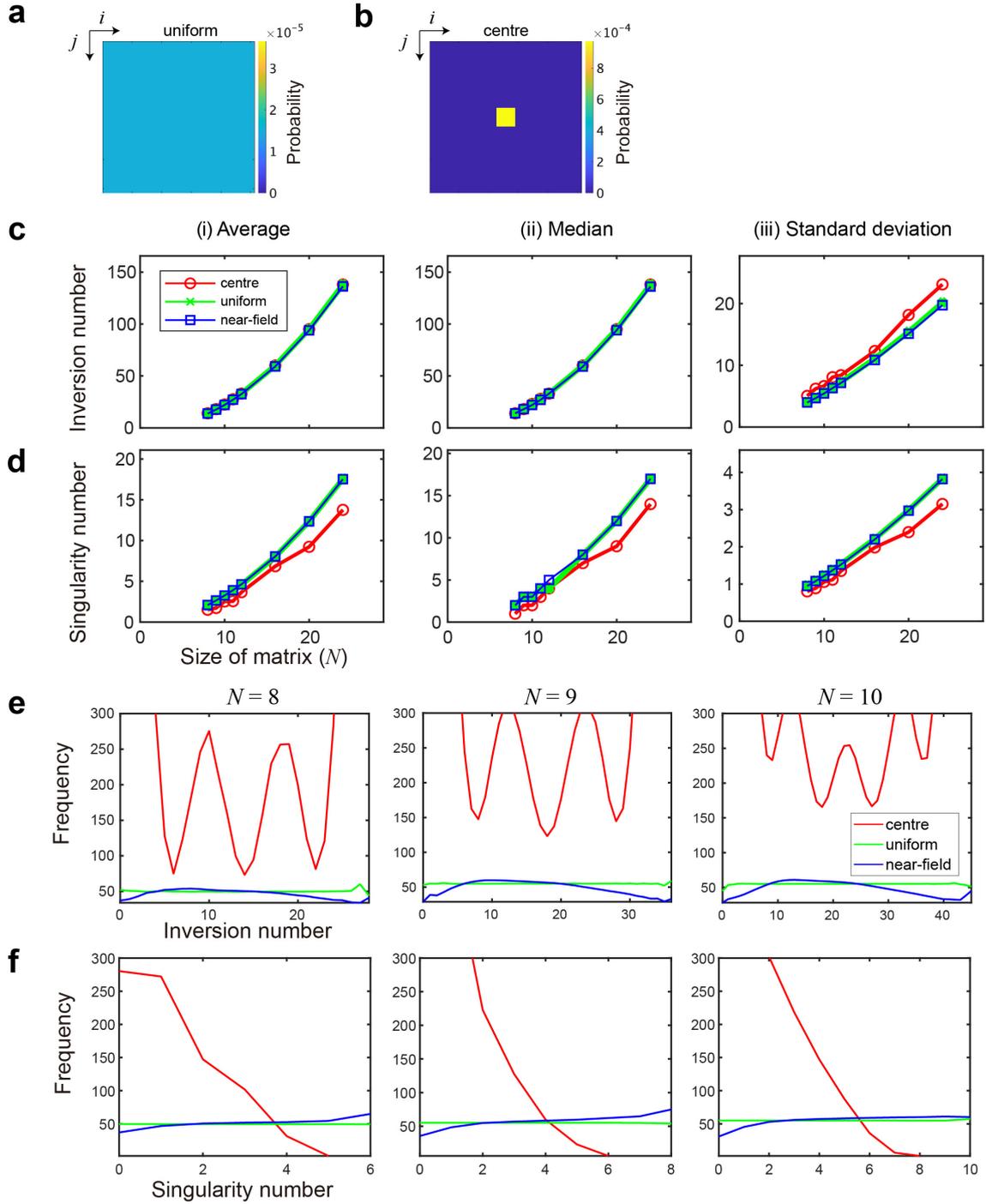

**Figure 6.** Characterization of Schubert polynomials generated by near-field pattern compared with (a) uniformly distributed and (b) spatially strongly skewed (centred) patterns. (c,d) Basic statistical analysis of (c) inversion number and (d) singularity number as a function of the size of the matrix ($N$). (i) Average, (ii) median, and (iii) standard deviations are examined where no evident differences among the sources are observed. (e,f) When the incidence is normalized by the number of unique matrices, evident source-dependencies are observed in both (e) inversion number and (f) singularity number.

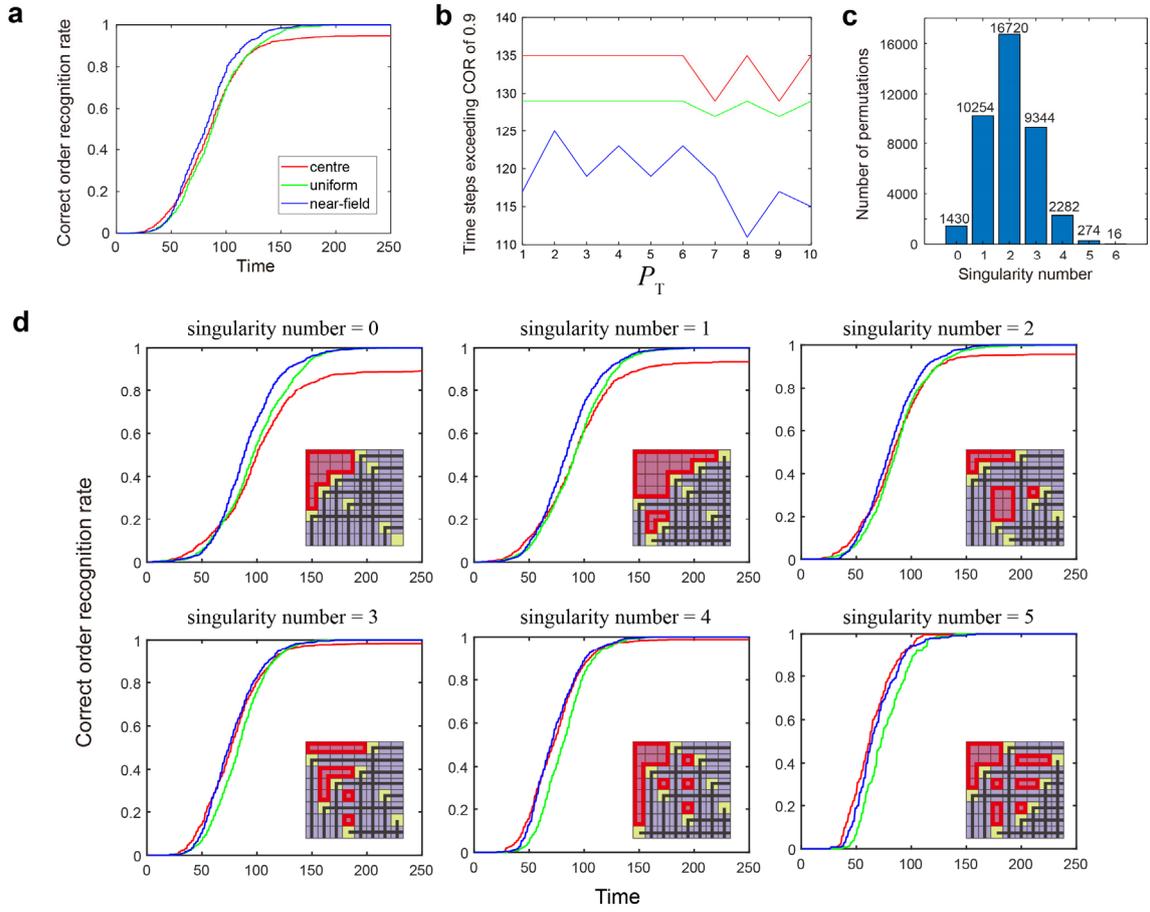

**Figure 7.** Order recognition performance evaluations with different source randomness using eight slot machine environments. (a) Time evolution of the correct order recognition (COR) rate for the near-field, uniform, and centre patterns. (b) The time steps exceeding the COR value of 0.9 as a function of the photon number threshold ($P_T$). (c) The breakdown of all permutations of $N = 8$ by the number of singularities. (d) Comparison of order recognition performances with problem settings having the same singularity value. When the singularity number is small (smaller than 2), near-field outperforms other sources, whereas, with the largest singularity, the centre pattern exhibits the promptest responses.

**Table A.1.** Schubert polynomials of $S_4$ and their inversion number and singularity number.

| Inversion number $l(\omega)$ | Permutation $\omega$ | Singularity number | Schubert polynomial $\mathfrak{S}_\omega = \partial_{\omega^{-1}\omega_0} \mathfrak{S}_{\omega_0}$ |
|---|---|---|---|
| 0 | 1234 | 0 | 1 |
| 1 | 2134 | 0 | $x_1$ |
|   | 1324 | 1 | $x_1 + x_2$ |
|   | 1243 | 1 | $x_1 + x_2 + x_3$ |
| 2 | 3124 | 0 | $x_1^2$ |
|   | 2341 | 0 | $x_1 x_2$ |
|   | 2143 | 1 | $x_1^2 + x_1 x_2 + x_1 x_3$ |
|   | 1423 | 1 | $x_1^2 + x_1 x_2 + x_2^2$ |
|   | 1342 | 1 | $x_1 x_2 + x_1 x_3 + x_2 x_3$ |
| 3 | 4123 | 0 | $x_1^3$ |
|   | 3214 | 0 | $x_1^2 x_2$ |
|   | 3142 | 1 | $x_1^2 x_2 + x_1^2 x_3$ |
|   | 2413 | 1 | $x_1^2 x_2 + x_1 x_2^2$ |
|   | 1432 | 1 | $x_1^2 x_2 + x_1^2 x_3 + x_1 x_2^2 + x_1 x_2 x_3 + x_2^2 x_3$ |
|   | 2341 | 0 | $x_1 x_2 x_3$ |
| 4 | 4213 | 0 | $x_1^3 x_2$ |
|   | 4132 | 1 | $x_1^3 x_2 + x_1^3 x_3$ |
|   | 3412 | 0 | $x_1^2 x_2^2$ |
|   | 3241 | 0 | $x_1^2 x_2 x_3$ |
|   | 2431 | 1 | $x_1^2 x_2 x_3 + x_1 x_2^2 x_3$ |
| 5 | 4312 | 0 | $x_1^3 x_2^2$ |
|   | 4231 | 0 | $x_1^3 x_2 x_3$ |
|   | 3421 | 0 | $x_1^2 x_2^2 x_3$ |
| 6 | 4321 | 0 | $x_1^3 x_2^2 x_3$ |

**Table A.2.** The number of generated matrices

| N  | Number of generated matrices | N!                    |
|----|------------------------------|-----------------------|
| 8  | $2 \times 10^6$              | $4.03 \times 10^4$    |
| 9  | $2 \times 10^7$              | $3.63 \times 10^5$    |
| 10 | $2 \times 10^8$              | $3.63 \times 10^6$    |
| 11 | $2 \times 10^9$              | $3.99 \times 10^7$    |
| 12 | $2 \times 10^9$              | $4.79 \times 10^8$    |
| 16 | $2 \times 10^9$              | $2.09 \times 10^{13}$ |
| 20 | $2 \times 10^9$              | $2.43 \times 10^{18}$ |
| 24 | $2 \times 10^9$              | $6.20 \times 10^{23}$ |